\documentclass[twocolumn]{aastex61}
\usepackage{graphicx}
\usepackage{tabularx}
\usepackage{amsmath,amssymb,bm}

\newcommand{\pd}[1]{\partial_{#1}}
\newcommand{\laplace}{\Delta}

\usepackage{multirow}

\newcommand{\AG}[1]{#1}

\submitjournal{ApJ}

\begin{document}

\title{Transport Properties of the Azimuthal Magnetorotational Instability}

\correspondingauthor{Anna Guseva}
\email{anna.guseva@zarm.uni-bremen.de}

\author{Anna Guseva}
\affiliation{University of Bremen, Center of Applied Space Technology and Microgravity (ZARM),  Bremen 28259, Germany}
\affiliation{Friedrich-Alexander University Erlangen-N\"urnberg,  Institute of Fluid Mechanics,  Erlangen 91058, Germany}

\author{Ashley P. Willis}
\affiliation{University of Sheffield, School of Mathematics and Statistics,  Sheffield S3 7RH, UK}

\author{Rainer Hollerbach}
\affiliation{University of Leeds, School of Mathematics,  Leeds LS2 9JT, UK}

\author{Marc Avila}
\affiliation{University of Bremen, Center of Applied Space Technology and Microgravity (ZARM),   Bremen 28259, Germany}
\affiliation{Friedrich-Alexander University Erlangen-N\"urnberg,  Institute of Fluid Mechanics,  Erlangen 91058, Germany}

\begin{abstract}

\AG{The magnetorotational instability (MRI) is thought to be a powerful source of  turbulence in Keplerian accretion disks. Motivated by recent laboratory experiments, we  study the MRI driven by an azimuthal magnetic field in an electrically conducting fluid sheared between two concentric rotating cylinders. By adjusting the rotation rates of the cylinders, we approximate  angular velocity profiles $\omega \propto r^{q}$. We perform direct numerical simulations of a steep profile close to the Rayleigh line $q \gtrsim -2 $ and a quasi-Keplerian profile $q \approx -3/2$ and cover wide ranges of Reynolds ($Re\le 4\cdot10^4$) and magnetic Prandtl numbers ($0\le Pm \le 1$). In the quasi-Keplerian case, the onset of instability depends on the magnetic Reynolds number, with $Rm_c \approx 50$, and angular momentum transport scales as  $\sqrt{Pm} Re^2$ in the turbulent regime. The ratio of Maxwell to Reynolds stresses is  set by $Rm$.  At the onset of instability both stresses have similar magnitude, whereas the Reynolds stress vanishes or becomes even negative as $Rm$ increases. For the profile close to the Rayleigh line, the instability shares these properties as long as $Pm\gtrsim0.1$, but exhibits a markedly different character if $Pm\rightarrow 0$, where the onset of instability is governed by the Reynolds number, with $Re_c \approx 1250$, transport is via Reynolds stresses and scales as $Re^2$. At intermediate $Pm=0.01$ we observe a continuous transition from one regime to the other, with a crossover at $Rm=\mathcal{O}(100)$. Our results give a comprehensive picture of angular momentum transport of the MRI with an imposed azimuthal field.} 
\end{abstract}


\section{Introduction}
\label{sec:intro}

The source of angular momentum transport  remained  the main question of accretion disk theory for years. Keplerian flows in accretion disks have radially decreasing angular velocity  $\omega \sim r^{-3/2}$ and are linearly stable according to the hydrodynamic Rayleigh criterion. However, the motion of gas in accretion disks cannot be laminar because viscous (molecular) outward transport is too slow for accretion to occur at the observed rates. Shakura \& Sunyaev suggested the presence of turbulent motion and parameterized momentum transport by an effective turbulent eddy-viscosity in their early $\alpha$-model \citep{shakura1973}. The origin of turbulence was unclear until  Balbus \& Hawley noted in 1991 that Keplerian flows of ionized gas can be destabilized by magnetic fields by the so-called magnetorotational instability \citep{balbus1991,balbus1998instability}. The MRI was first described by  \citet{velikhov1959stability} and \citet{chandrasekhar1961}, who investigated the stability of electrically conducting fluids sheared between two concentric cylinders (Taylor-Couette flow) and subjected to an axial (poloidal) magnetic field. The MRI operates if the angular velocity decreases outwards and angular momentum increases, which is  the case in Keplerian flows. 

Since the seminal work of  \citet{balbus1991} there has been considerable interest in the MRI from the theoretical, numerical and experimental points of view. The action of poloidal fields, which can be generated by the accreting object in the center of the disk or advected from outside, is well-studied. Weak poloidal magnetic fields lead to the amplification of axisymmetric disturbances and give rise to self-sustained turbulence in nonlinear simulations \citep{balbus1991,hawley1995local,stone1996three}. This MRI turbulence was found to significantly enhance angular momentum transport via Maxwell stresses, which were several times larger than Reynolds stresses. However, these early works did not take  account of viscosity and magnetic resistivity (ideal MHD). \citet{lesur2007impact} considered  non-ideal fluids and showed a power-law dependence of the transport coefficient of the form $\alpha \sim Pm^{\gamma}$, with $\gamma \in (0.25,0.5)$, for  the explored range of magnetic Prandtl numbers $Pm \in [0.12, 8]$ and Reynolds numbers $Re \in [200, 6400]$.

It is important to note that most nonlinear simulations of the MRI have been performed using the shearing sheet approximation, which consists of a local model of an accretion disk \citep{hawley1995local}. In this approximation, the equations are solved in a rotating frame in Cartesian geometry, with the rotation given by the linearization of the Keplerian law at a  radial point in the disk. Periodic boundary conditions are assumed in all three directions, and radial shear is introduced by means of a coordinate transformation. These boundary conditions determine the geometry of the modes observed in the simulations and their saturation in the nonlinear regime \citep{regev2008viability}. In addition, most simulations of shearing boxes neither resolve all flow scales nor implement subgrid models that capture the impact of small flow scales on the larger scales.

\citet{ji2001magnetorotational} and \citet{rudiger2001mhd} suggested the study of the MRI in an electrically conducting fluid sheared between two concentric cylinders, exactly as considered originally by  \citet{velikhov1959stability} and \citet{chandrasekhar1961}. By appropriately choosing the rotation-ratio of the cylinders, velocity profiles of the general form  $\omega(r) \sim r^q$, including $q=-1.5$ for Keplerian rotation, can be well approximated experimentally at very large Reynolds numbers \citep{edlund2015reynolds,lopez2017boundary}. For an imposed axial magnetic field, \citet{gellert2012angular} found the transport coefficient $\alpha$ to be independent of magnetic Reynolds $Rm$ and magnetic Prandtl $Pm$ numbers, only scaling linearly with the Lundquist number $S$ of the axial magnetic field.  However, the critical parameter values $Rm\sim O(10)$, $S\sim O(3)$ are challenging to achieve experimentally because of the low values of $Pm\in[10^{-6},10^{-5}]$ for liquid metals \AG{and therefore extremely high critical Reynolds numbers $Re = Rm/Pm \ge O(10^{6})$}. \AG{While signatures of this standard MRI were detected experimentally in the form of damped magnetocoriolis waves \citep{nornberg2010observation}, unstable magnetocoriolis waves (giving rise to the MRI) have not been reported in the literature so far}.

The MRI can also be triggered by toroidal magnetic fields provided that these are neither too weak nor too strong \citep{balbus1992powerful,ogilvie1996non}. Interest in this ``azimuthal'' MRI (AMRI) increased further following the linear stability analysis of  \citet{hollerbach2010nonaxisymmetric}, who found that for steep velocity profiles close to the Rayleigh line ($q\gtrsim-2$) the AMRI is governed by the Reynolds $Re$ and Hartmann  $Ha$ numbers, instead of $Rm$ and $S$. The two sets of parameters are related by $Rm=Pm\,Re$ and $S=\sqrt{Pm}\,Ha$. \AG{For $Pm \to 0$ this inductionless version of MRI continues to exist even for $S$, $Rm \to 0$, as long as $Re \sim O(10^3)$ and $Ha\sim O(10^2)$. It takes the form of an inertial wave destabilized by the magnetic field via the Lorentz force, and hence is a magnetohydrodynamic rather than hydrodynamic instability 
\citep[see][for an extended discussion and connection to the standard MRI via helical fields]{Kirillov2010}. }

\citet{seilmayer2014experimental} reported the experimental observation of the predicted non-axisymmetric AMRI modes in Taylor--Couette flow \AG{with $q=-1.94$}. Because of the strong currents of nearly 20kA needed to generate the required azimuthal field, measurements could only be conducted close to the stability boundary. Using direct numerical simulations \citet{guseva2015transition} probed deep into the nonlinear regime and computed the angular momentum transport of the AMRI. Despite the highly turbulent nature of the flow, for $Pm=1.4\cdot10^{-6}$ (InGaSn alloy) the angular momentum transport was found to be barely faster than in laminar flow. Recently, \citet{rudiger2015angular} examined the effective viscosity $\nu_t$ for the three relevant rotation rates: close to the Rayleigh line ($q \sim -2$), quasi-Keplerian ($q \sim -1.5$) and galactic ($q \sim -1$), in the range of $Pm \in [10^{-1},1]$ and $Re \in [2\cdot 10^2,2 \cdot 10^3]$. They suggested  a scaling of the dimensionless  effective viscosity as $\nu_t/\nu=\sqrt{Pm}Re$, with Maxwell stresses dominating for large $Pm \geq 0.5$. However, in the range of parameters investigated in \citet{rudiger2015angular},  AMRI turbulence \AG{does not yet clearly exhibit asymptotic scaling} and the transition between low-$Pm$ and high-$Pm$ instability and transport properties at low and moderate $Pm$ remain unclear.

In this paper, we  first revisit the linear stability analysis of the AMRI by considering various rotation laws, over a range of magnetic Prandtl, Hartman and Reynolds numbers. We show the scalings determining the existence of the instability as a function of these parameters. Second, using direct numerical simulations, we compute angular momentum transport in the system for $Re$ up to $4\cdot10^4$ and $Pm\in [0,1]$.  
Our results give a comprehensive picture of turbulent transport via the  AMRI.

\section{Model}

\subsection{Governing equations and parameters}
\label{sec:equations}

We consider an incompressible viscous electrically conducting fluid sheared between two rotating cylinders \AG{of inner and outer radii $r_i$ and $r_o$}. The velocity $\bm{u}$ and magnetic field $\bm{b}$ are determined by the coupled Navier--Stokes and \AG{induction} equations:
\begin{equation}\label{eq:NSeq}
   (\pd{t} + \bm{u}\cdot\nabla) \bm{u} = - \frac{1}{\rho}\nabla p + \nu \laplace \bm{u} + \frac{1}{\mu_0 \rho} (\nabla \times \bm{b}) \times \bm{b},
\end{equation}
\begin{equation}\label{eq:Ieq}
( \pd{t} - \lambda \laplace ) \bm{b}  = \nabla \times (\bm{u} \times \bm{b}),
\end{equation}
together with $\nabla\cdot{\bm{u}}=\nabla\cdot{\bm{b}}=0$. \AG{Here $p$ is the fluid pressure, $\rho$, $\nu$, and $\lambda$ are the constant density, kinematic viscosity and magnetic diffusivity of the fluid, respectively. The Navier--Stokes and induction equations \eqref{eq:NSeq}--\eqref{eq:Ieq} are formulated in cylindrical coordinates $(r, \phi, z)$ with periodic boundary conditions in the axial and azimuthal directions. 
Insulating boundary conditions are imposed for $\bm{b}$.  No-slip boundary conditions are imposed for $\bm{u}$, which for the azimuthal velocity read $u_{\phi} (r_i) = \omega_i r_i$, $u_{\phi} (r_o) = \omega_o r_o$, where $\omega_i$ ($\omega_o$) is the angular speed of the inner (outer) cylinder. The  laminar  angular velocity profile is}
\begin{equation}\label{eq:TCFprofile}
\omega_\text{lam}=\frac{\omega_i}{1- \eta^2} \left[\left(\mu - \eta^2 \right)+ r_i^2 \left(1-\mu \right) \frac{1}{r^2}\right].
\end{equation}\\
 
In this work we take the radius ratio to be $\eta=r_i/r_o=0.5$, and the rotation ratio $\mu=\omega_o/\omega_i$ is varied to approximate different rotation laws. Close to the Rayleigh line we focus on $\mu=0.26$, and for quasi-Keplerian rotation we take $\mu=0.35$, which  yield $q=-1.94$ and $q=-1.48$ respectively, based on the average shear. The remaining dimensionless parameters of the problem are the Hartmann \AG{$Ha={B_0}d/(\sqrt{\mu_0 \rho \nu \lambda})$}, Reynolds $Re=\omega_i r_i d /\nu$ and magnetic Reynolds  $Rm=\omega_i r_i d /\lambda$ numbers, \AG{where $d=r_o-r_i$ is the gap between cylinders}. $Re$ and $Rm$ are connected through the magnetic Prandtl number $Pm=Rm/Re=\nu/\lambda$. 

\subsection{Numerical methods}

For the linear stability analysis of the laminar flow in \S\ref{sec:linear}, the spectral eigenvalue solver of \citet{hollerbach2010nonaxisymmetric} was employed. The fully coupled nonlinear Navier--Stokes and induction equations \eqref{eq:NSeq}--\eqref{eq:Ieq} were discretized with high-order finite-differences in the radial direction and the Fourier pseudospectral method in the axial and azimuthal directions. The time discretization is based on the implicit Crank--Nicolson method and is of second order. Details of our numerical method, implementation and tests can be found in \citet{guseva2015transition}. The numerical resolution was chosen so that our simulations were fully resolved; it reached $N=480$ finite-difference points in the radial direction and $720$ ($K=\pm 360$) and $560$ ($M=\pm280$) Fourier modes in the axial and azimuthal directions. The aspect-ratio in the axial direction was fixed to $L_z=h/d=1.4$ (high $Re$) or $12.6$ (low $Re$), where $h$ is the length of the cylinders. 

\subsection{Angular velocity current}
\label{sec:vel_current}

Frequently a net loss (gain) of angular momentum on the inner (outer) cylinder is measured in experiments as torque \citep{paoletti2011angular,wendt1933turbulent}. Dimensionless laminar torque $G_{lam}$ can be explicitly calculated from the radial derivative of azimuthal velocity profile \eqref{eq:TCFprofile},
\begin{equation}\label{eq:Glam}
 G_\text{lam} = -\frac{2 \pi}{\nu} r^3 \partial_r (\omega_\text{lam}), 
\end{equation}
 \AG{or equivalently,}
\begin{equation}\label{eq:Glam_nondim}
G_\text{lam} = \frac{4 \pi \eta |\mu -1|}{ (1- \eta)^2 (1+\eta)} Re.
\end{equation}
In a statistically steady state, the time-averaged torques on the inner and outer cylinders are equal in magnitude.

\citet{eckhardt2007torque} derived a conservation equation for the current $J^\omega$ of the angular velocity $\omega=u_\phi/r$ in hydrodynamic Taylor--Couette flow. In this work we extend this to the magnetohydrodynamic case as follows. Defining
\begin{equation}
p' = \frac{p}{\rho} +   \frac{1}{\mu_0 \rho} \frac{\bm{b}^2}{2},
\end{equation}
the $\phi$-component of the Navier--Stokes equation becomes (\ref{eq:u_phi}). Averaging over time
and a co-axial cylindrical surface of area $A(r) = 2 \pi r h$  yields (\ref{eq:average}). Using the divergence-free condition for $\bm{u}$ and $\bm{b}$, and multiplying by $r^2$, we finally obtain equation (\ref{eq:stress}).
\onecolumngrid
\begin{eqnarray}\label{eq:u_phi}
\pd{t} u_{\phi} = - (\bm{u} \cdot \nabla) u_{\phi} - \frac{u_r u_{\phi}}{r} - \frac{1}{r} \pd{\phi} (p') + \nu \Big(\laplace u_{\phi} - \frac{u_\phi}{r^2} + \frac{2}{r^2} \pd{\phi} u_r \Big) 
+ \frac{1}{\mu_0 \rho} \Big( (\bm{b} \cdot \nabla) b_\phi + \frac{b_r b_{\phi}}{r} \Big)
\end{eqnarray}
\begin{eqnarray}\label{eq:average}
0 = \Big\langle \Big(- u_r \pd{r} u_\phi - u_z \pd{z} u_\phi - \frac{u_r u_\phi}{r}\Big)
 +  \nu \Big( \frac{1}{r} \pd{r} (r \pd{r} u_\phi) - \frac{u_\phi}{r^2}  \Big)
+ \frac{1}{\mu_0 \rho}\Big(b_r \pd{r} b_\phi + b_z \pd{z} b_\phi + \frac{b_r b_\phi}{r} \Big) \Big\rangle_{A,t}
\end{eqnarray}
\begin{eqnarray}\label{eq:stress}
\pd{r}(J^\omega)=0\qquad\text{where}\qquad J^\omega\equiv
r^3 \Big(\Big\langle u_r \omega \Big\rangle_{A,t} - \nu  \pd{r} \Big\langle \omega \Big\rangle_{A,t} -  \frac{1}{\mu_0 \rho}\Big\langle \frac{b_r b_\phi}{r} \Big\rangle_{A,t} \Big)
\end{eqnarray}
\twocolumngrid
The first term in equation \eqref{eq:stress} represents a Reynolds stress, whereas the second and third terms represent viscous and Maxwell stresses, respectively. For the steady rotation $J^\omega$  is conserved [$\partial_r (J^\omega)=0$, see equation (\ref{eq:stress})]. The constant $J^\omega$ can be interpreted as the conserved transverse current of azimuthal motion transporting $\omega(r,\phi,z,t)$ in the radial direction. Its unit is [$J^\omega$] = m$^4$ s$^{-2}$ = [$\nu$]$^2$.    The  angular velocity current $J^\omega$ is closely related to the dimensionless torque on the cylinders \citep{eckhardt2007torque}:
\begin{equation}\label{eq:stress2torque}
G= 2 \pi \nu^{-2} J^\omega.
\end{equation} 
\AG{In the case of laminar flow $ \langle \omega \rangle_{A,t} =\omega_\text{lam}$ 
the first and the third term in Equation \eqref{eq:stress} are zero and the laminar angular velocity current is defined by 
\begin{equation}\label{eq:lam_jw}
 J^\omega_\text{lam} = - \nu r^3 \partial_r \omega_\text{lam},
\end{equation} 
so that \eqref{eq:stress2torque} and \eqref{eq:Glam} coincide.
}

\subsection{Effective viscosity}

\label{sec:turb_visc}

For the case of turbulent flow, in analogy to \eqref{eq:lam_jw}
we model angular velocity current with the mean angular velocity:
\begin{equation}\label{eq:turb_model}
 J^\omega = - {\nu}_\text{eff} r^3 \partial_r \langle \omega \rangle,
\end{equation}
where the effective viscosity $\nu_\text{eff}$ is parameterized with the mean angular velocity and the size of the gap between cylinders
\begin{equation}\label{eq:nu_t_alpha}
\nu_\text{eff}=\alpha_\text{eff} \langle \omega \rangle d^2.
\end{equation}
Substituting \eqref{eq:nu_t_alpha} into \eqref{eq:turb_model} results in the following for  the parameter $\alpha_\text{eff}$:
\begin{equation}\label{eq:alpha_deriv}
\alpha_\text{eff} = - \frac{J^\omega}{\langle \omega \rangle d^2 r^3 \partial_r \langle \omega \rangle} = - \frac{J^\omega}{\langle \omega \rangle^2 d^2 r^2 q} \frac{\nu^2}{\nu^2}= \frac{J^\omega }{|q| \nu^2 Re_*^2}.
\end{equation}
Here $q=\partial \ln \langle \omega \rangle / \partial \ln r$ and $Re_*$ is a Reynolds number based on the average angular velocity. Estimation of $Re_*$ based on $ \langle \omega \rangle \approx \left( \omega_i + \omega_o \right)/2$ and the midgap radius  $r=\left(r_i+r_o\right)/2$ gives $Re_*^2=1.026 Re^2$. Recalling \eqref{eq:stress2torque} we finally obtain:
\begin{equation}\label{eq:alpha_fin}
\alpha_\text{eff}=\frac{1}{2 \pi |q|} \frac{G}{Re^2}.
\end{equation}

Protostellar-disk accretion rates indicate $\alpha_\text{eff} \geq10^{-3}$ \citep{hartmann1998accretion}.

\section{Stability analysis}

\subsection{Linear stability analysis}\label{sec:linear}

\begin{figure}
\centering
\begin{tabular}{c}
    (a)\\
     \includegraphics[width=0.9\linewidth]{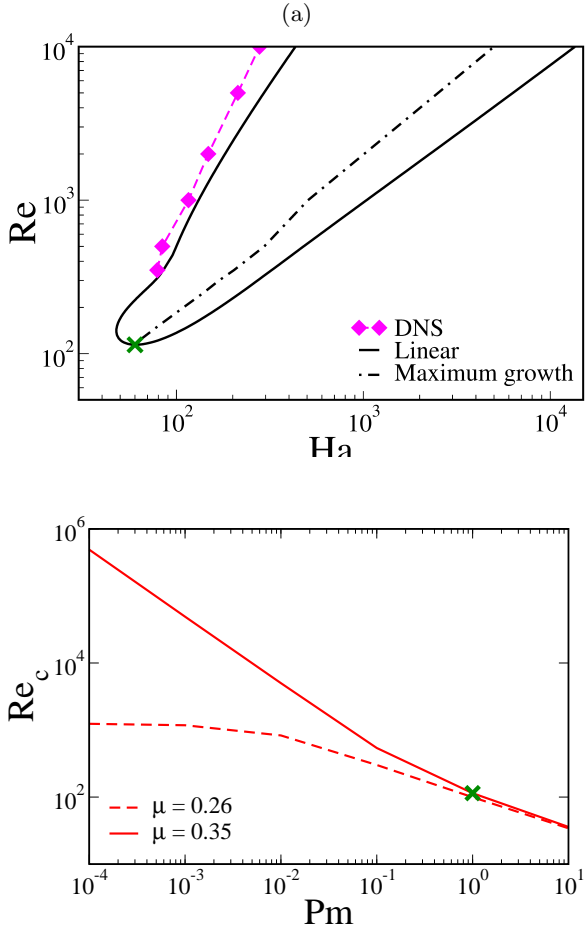}\\
     (b)\\
    \includegraphics[width=0.9\linewidth]{Re_cr_Pm_paper_2.eps}
     \end{tabular}
  \caption{\small (a) Neutral curve from the linear stability analysis (black solid line), nonlinear stability border (magenta diamonds) and maximum growth rate of the instability at fixed $Re$ as a function of $Ha$ (black dashed-dotted line) for $\mu=0.35$ and $Pm=1$. The green cross marks the minimum of the neutral curve $Re_\text{c}=Re(Ha)$, which is the lowest Reynolds number at which the laminar flow can be destabilized. (b) $Re_\text{c}$ as a function of $Pm$ for $\mu=0.26$ (dashed line) and $\mu=0.35$ (solid line). The green cross indicates the value for $\mu=0.35$ and $Pm=1$, as in (a).
  } 
  \label{fig:linear_scaling}
\end{figure}

We begin by studying stability of the laminar flow to infinitesimal disturbances. For the curvature considered here, $\eta=0.5$, the dominant AMRI mode is non-axisymmetric with azimuthal wavenumber $m=1$. \AG{At fixed parameters $(Ha, Re, Pm, \mu)$, the axial wavenumber $k$ was varied to determine the maximum perturbation growth rate.  The flow} becomes unstable only after $Re$ exceeds a critical threshold and remains unstable thereafter only in a certain range of $Ha$, i.e.~for neither too weak nor too strong magnetic fields \citep{hollerbach2010nonaxisymmetric}. The black solid curve in Fig.~\ref{fig:linear_scaling}a shows the neutral stability curve of the AMRI for $Pm=1$ and $\mu=0.35$. In the region enclosed by this curve the laminar flow is linearly unstable. The growth rate of the instability is maximized (for $Re$=const) along the black dashed-dotted line. The curve starts at $Ha=60$ and $Re_\text{c}=114$, which is the minimum Reynolds number at which the instability occurs. Figure~\ref{fig:linear_scaling}b shows the dependence of $Re_\text{c}$ as a function of $Pm$ for  $\mu=0.35$ and $\mu=0.26$. At large $Pm>1$ the two curves collapse, indicating that the AMRI becomes insensitive to rotation profile. In fact for $Pm\gtrsim 10$ the AMRI occurs at lower $Re$ than the Rayleigh (centrifugal) instability at all $\mu$. As $Pm$ decreases the two curves gradually depart from each other. For quasi-Keplerian rotation ($\mu=0.35$) and $Pm<0.1$,  $Re_\text{c}$ grows inversely proportional to $Pm$, and so the onset of instability occurs at a constant magnetic Reynolds number $Rm_c=Re_c\,Pm\approx 50$. By contrast, for $\mu=0.26$ the onset of instability occurs at a constant hydrodynamic Reynolds number $Re_c \approx 1250$. 

\begin{figure}
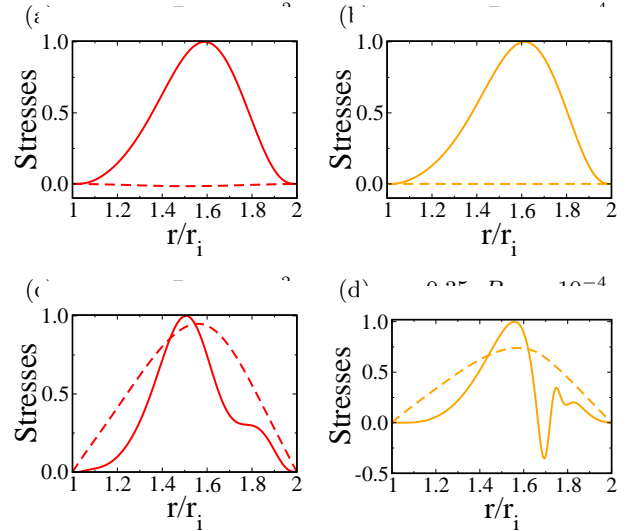

\centering
\begin{tabular}{ccc}
(a) $\mu=0.26$,  $Pm=10^{-2}$ & (b) $\mu=0.26$, $Pm=10^{-4}$\\
     \includegraphics[width=0.45\linewidth]{Pm1d-2_mu26_bold.eps} &
    \includegraphics[width=0.45\linewidth]{Pm1d-4_mu26_bold.eps} \\
    (c) $\mu=0.35$,  $Pm=10^{-2}$& (d) $\mu=0.35$,  $Pm=10^{-4}$\\
     \includegraphics[width=0.45\linewidth]{Pm1d-2_mu35_bold.eps} &
    \includegraphics[width=0.45\linewidth]{Pm1d-4_mu35_bold.eps} \\
 \end{tabular}
\AG{  \caption{\small Reynolds (solid) and Maxwell (dashed) stresses of the eigenmodes near the onset of instability for rotation close to the Rayleigh line  (first row) and quasi-Keplerian rotation (second row). The Reynolds numbers are  
(a) $Re=840$ ($Rm=8.4$), (b) $Re=1300$ ($Rm=0.13$), (c) $Re=5000$ ($Rm=50$), (d) $Re=5\cdot 10^5$, ($Rm=50$). } } 
  \label{fig:linear_stress}
\end{figure}

\AG{To shed more light on the difference in behavior in the $Pm \rightarrow 0$ limit, we analyzed Maxwell and Reynolds stresses for the critical eigenmodes. The radial distributions of stresses for $\mu=0.26$ and $\mu=0.35$ at low $Pm$ are shown in the first and second row of Fig.~\ref{fig:linear_stress}, respectively.  The ratio of Maxwell to  Reynolds stresses, each integrated over the radius $r$, for  $\mu=0.35$ is about $1.5$ for both  $Pm = 10^{-2}$ and  $10^{-4}$, whereas for $\mu=0.26$ it decreases from $0.018$ to $0.00027$ as $Pm$ is reduced from $10^{-2}$ to $10^{-4}$. This suggests that the relative importance of stresses is essentially set by the magnetic Reynolds number. }

\AG{Overall, the linear analysis of this section highlights the markedly different character of the instability at low $Pm$ for rotation near the Rayleigh line, with $Re_c \approx 1250$ (magnetically destabilized inertial wave with transport via Reynolds stresses), and quasi-Keplerian rotation, with $Rm_c \approx 50$ (unstable magnetocoriolis wave with transport via Maxwell and Reynolds stresses).}

\subsection{Nonlinear stability analysis}

Close to the onset of instability the AMRI is a rotating wave in the azimuthal direction~\citep{hollerbach2010nonaxisymmetric}, whereas in the axial direction it can be a standing wave (SW) or a traveling wave (TW) \citep{knobloch1996symmetry}. For $Pm=1.4 \cdot 10^{-6}$ a SW is realized \citep{guseva2015transition}, whereas at $Pm=1$ the AMRI manifests itself as a TW, thereby breaking the axial reflection symmetry \citep{guseva2016azimuthal}. Close to $Re_c$ the bifurcation is found to be supercritical in all cases. However, at large $Pm=1$ and $Re>3 \cdot 10^2$ the nonlinear AMRI pattern survives outside the left stability border (magenta diamonds in Fig.~\ref{fig:linear_scaling}a), indicating a subcritical bifurcation for both $\mu=0.26$ and $\mu=0.35$. As a result, for $Pm=1$ the left stability border widens faster than expected from the linear estimate.

\section{Angular momentum transport}\label{sec:nonlinear:transport}

\begin{figure}
\centering
    \includegraphics[width=0.9\linewidth]{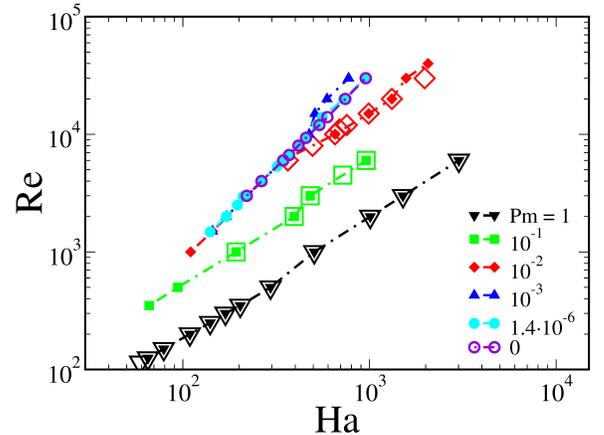} 
  \caption{\small Filled symbols show the parameter values at which our DNS were performed ($\mu=0.26$). They follow one-dimensional curves in $(Re,Ha)$-space, corresponding to the maximum growth rate lines of the linear stability analysis (see Fig. \ref{fig:linear_scaling}a). Data for quasi-Keplerian rotation $\mu=0.35$ at \AG{$Pm=1$,  $10^{-1}$ and $10^{-2}$ } are shown as empty symbols of the same color. The violet empty circles correspond to DNS of the inductionless limit ($Pm=0$) at $\mu=0.26$. 
   } 
  \label{fig:param}
\end{figure}

We performed DNS spanning a wide range in $Pm \in [0, 1]$ and $Re$ up to $4 \cdot 10^4$ and computed the torque $G$ in order to quantify the scaling of angular momentum transport via the AMRI. Each filled symbol in Fig.~\ref{fig:param} marks a simulation in the parameter space $(Re,Ha)$, with $\mu=0.26$. The simulations with $\mu=0.35$ follow the same path as for $\mu=0.26$ and are shown as empty symbols of the same color.  Because of the cost of varying both $Ha$ and $Re$, we followed one-dimensional paths in parameter space (dashed-dotted lines connecting the symbols) that correspond to the maximum growth rate lines of the linear analysis (depicted by the dashed-dotted line in Fig.~\ref{fig:linear_scaling}a for $Pm=1$ and $\mu=0.35$). Note that \citet{guseva2015transition,guseva2016azimuthal} showed that the maximum growth rate of the linear analysis correlates very well with the maximum of the transport for $\mu=0.26$ at low and high $Pm$. \cite{mamatsashvili2017} have observed the same correlation for the helical MRI. Hence the results presented in the following can be seen as an upper bound on the angular momentum transport. 

The value of $Re_c$ depends strongly on $\mu$ and $Pm$ as shown in Fig.~\ref{fig:linear_scaling}b. Hence comparing the scaling of $G$ with $Re$ at different $Pm$ and $\mu$ is not straightforward. Moreover,  \citet{guseva2015transition,guseva2016azimuthal} found that at low $Pm$, close to the Rayleigh line, the turbulence arising from the AMRI is not  efficient in transporting momentum. At $Pm=10^{-6}$ and \AG{$Re\lesssim3\cdot10^4$},  molecular viscosity is responsible for a significant portion of the momentum transport. As a consequence, the laminar contribution to the torque can obscure the scaling of the turbulent contribution, which will obviously dominate at the asymptotically large $Re$ of interest. To enable a representation more useful for extrapolations toward large $Re$, in this section we quantify transport by showing the reduced torque ($G/G_\text{lam} -1$), which is proportional to the turbulent viscosity, as a function of the relative Reynolds number $Re' = Re-Re_\text{c}$. 

\subsection{Close to the Rayleigh line}

\begin{figure}
\centering
(a)\\
    \includegraphics[width=0.95\linewidth]{torque_scale_paper_mu26.eps} \\
 (b)\\
    \includegraphics[width=0.95\linewidth]{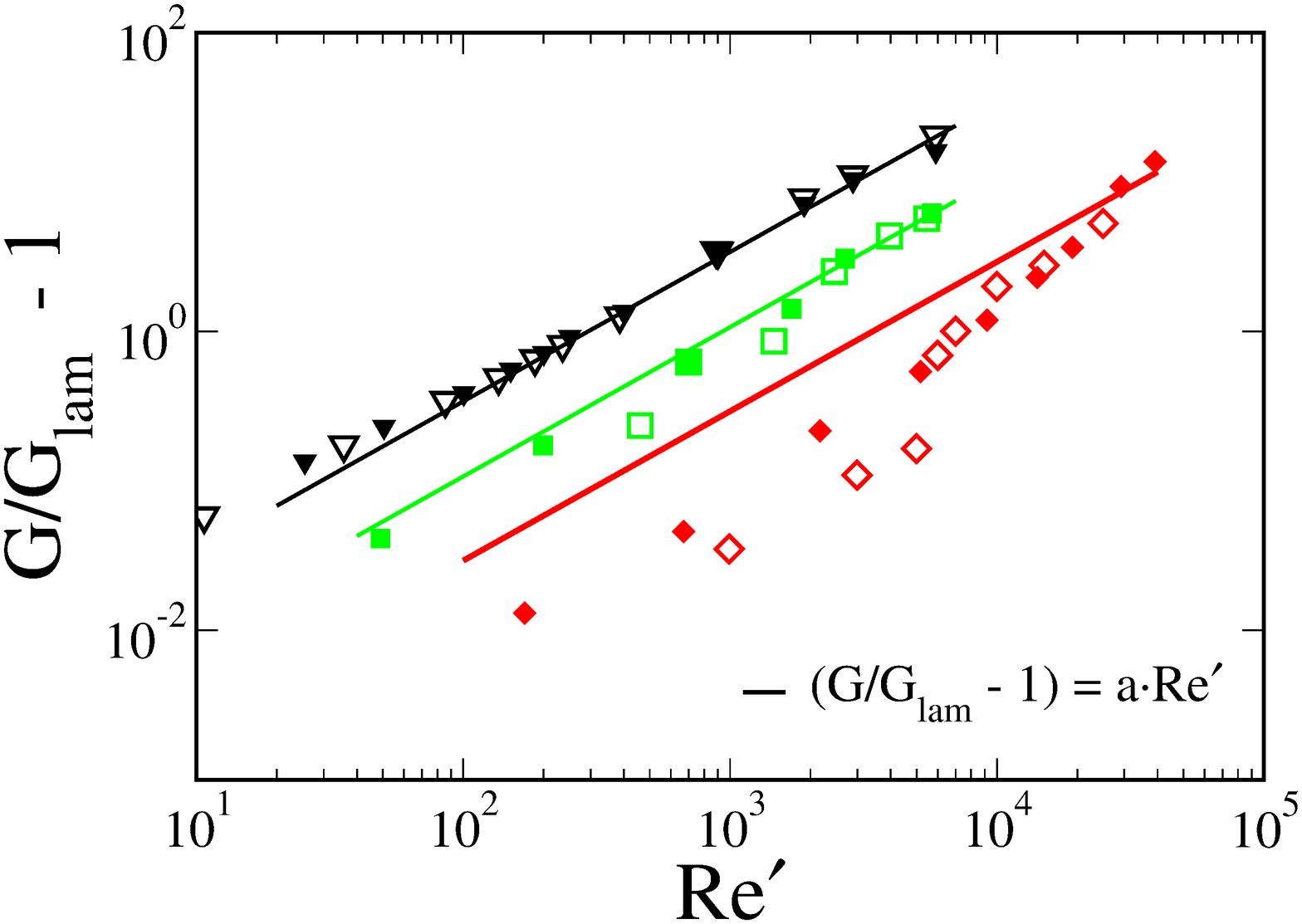} \\
 (c)\\
  \includegraphics[width=0.9\linewidth]{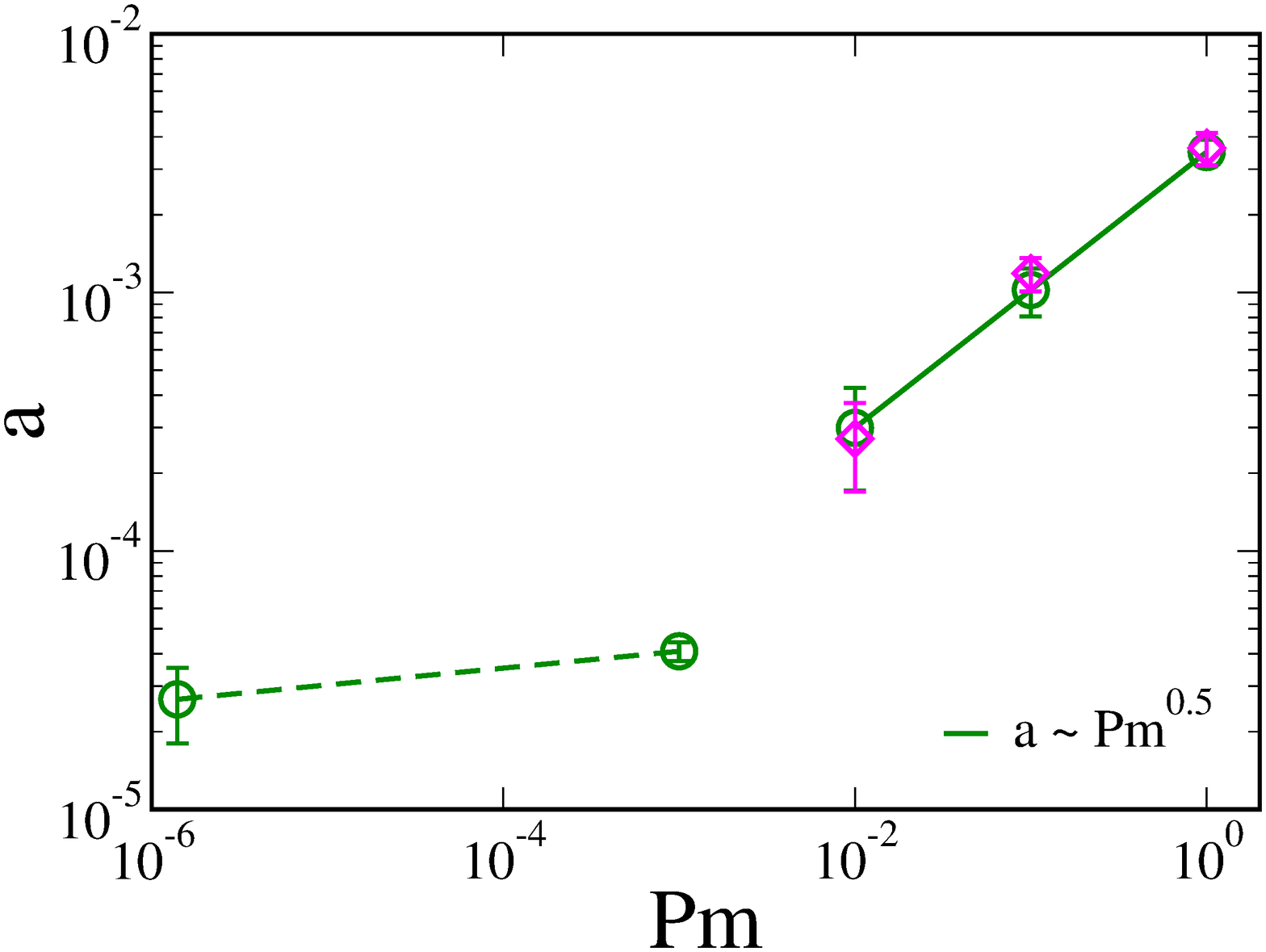} \\
  \caption{\small (a) Normalized turbulent torque $(G/G_\text{lam} -1)$ for $\mu=0.26$ as a function of modified Reynolds number ($Re'=Re-Re_{cr}$).  $Pm=1$ - black triangles, $Pm=10^{-1}$ - green squares, $Pm=10^{-2}$ - red diamonds, $Pm=10^{-3}$ - blue triangles, and $Pm=1.4 \cdot 10^{-6}$ - cyan circles, $Pm=0$ - violet empty circles.  For each $Pm$ a line $(G/G_\text{lam} -1)=a Re'$ is fitted.  (b) \AG{Comparison of the quasi-Keplerian rotation $\mu=0.35$ (empty symbols) to $\mu=0.26$ (filled symbols). Same color code as in the Fig. \ref{fig:torque}a.} 
 (c) Average scaling factor $a$ as a function of Pm. \AG{Dark green - $\mu=0.26$, magenta - $\mu=0.35$.}} 
  \label{fig:torque}
\end{figure}

Figure \ref{fig:torque}a shows that for $\mu=0.26$ the reduced torque scales linearly with $Re'$, whereas the dependence on $Pm$ is not straightforward.  Lines of $(G/G_\text{lam} -1)=a\,Re'$, where the pre-factor $a$ is a function of $Pm$, provide good fits to all data sets.  The pre-factors $a$ and respective errors were calculated based on the average of  local fits to the data using stencils of 3 up to 5 points. The dependence of the pre-factor $a$ on $Pm$ is shown in  Fig~\ref{fig:torque}c. 
For $Pm \geq 10^{-2}$, $a \propto Pm^{0.53}$, supporting the $\sqrt{Pm}\,Re$ scaling proposed by \citet{rudiger2015angular}, whereas for $Pm \leq 10^{-3}$,  $a$ saturates to a small but constant value independent of $Pm$. We verified this by performing DNS in the inductionless limit ($Pm=0$, violet empty circles in \ref{fig:torque}a) and the results are in fact indistinguishable from  $Pm=1.4 \cdot 10^{-6}$ ($a(Pm=0)=2.3\pm 0.6\cdot 10^{-5}, a(Pm=1.4 \cdot 10^{-6})=2.7\pm 0.8 \cdot 10^{-5}$). This supports the hypothesis that in the limit of very small magnetic Prandtl numbers $Pm \to 0$ the turbulent angular momentum transport triggered by AMRI turbulence depends only on $Re$, and this dependence is linear.

The case of $Pm=10^{-2}$ requires special attention. \AG{Close to the onset of instability the scaling factor $a$ tends to the low-$Pm$ values and  the $\sqrt{Pm}\, Re$-scaling is approached only at high $Re$. This is most likely connected to the change in the dominant mode at $Re\approx 5\cdot10^3$ ($Rm\approx 50$), illustrated by the change of slope of the maximum growth rate in Fig.~\ref{fig:param}. }  Thus, the quality of the  linear fit for $Pm=10^{-2}$ is not very good, and the error bar for $a$ on Figure \ref{fig:torque} is the largest. Still the average value of the pre-factor $a$ approximates  well the local fits of the high-$Re$ part of the curve, and can be taken as an estimate.

\subsection{Quasi-Keplerian rotation}

Although the case of $\mu=0.26$ allows us to span a broad range of relevant $Pm$, the astrophysically most relevant profile is the quasi-Keplerian one ($\mu=0.35$). To compare the torque of the two rotation rates, we performed simulations at $\mu=0.35$ and $Pm=1$, $10^{-1}$ and $10^{-2}$  where $Re_\text{c}$ is low enough for DNS to be feasible (Fig. \ref{fig:linear_scaling}b). \AG{The results of this comparison are presented in Fig.~\ref{fig:torque}b.} Unlike in \citet{rudiger2015angular}, we do not observe that the torque is lower for the $\mu=0.35$ profile. The lower values of torque at low $Re$ are because of the later onset of instability at $\mu=0.35$ and converge toward the values for the $\mu=0.26$ case as the turbulence develops further. Hence, once the dependence of $Re_\text{c}$ on $\mu$ is taken into account, the torque scales identically for both rotation profiles, as demonstrated in Figure \ref{fig:torque}c. \AG{Because for $\mu=0.35$ the onset of instability occurs at $Rm_c\approx 50$, studying $Pm<10^{-2}$ becomes numerically unfeasible. Hence it cannot be directly tested whether a transition from the $\sqrt{Pm}\, Re$-scaling to the pure $Re$-scaling, as observed close to the Rayleigh line,  occurs also in the quasi-Keplerian case.}

\section{Analysis of transport mechanisms}

\AG{The analysis of the Maxwell and Reynolds stresses of the linear eigenmodes at low $Pm$ shown in \S\ref{sec:linear} suggests that for $\mu=0.35$ Reynolds and Maxwell stresses are relevant, whereas for $\mu=0.26$ only Reynolds stresses play a role. In this section we analyze  the dependence of the stress contributions for the data from the nonlinear simulations shown in Fig.~\ref{fig:torque}.} In nonlinear simulations of the fully coupled Navier--Stokes and \AG{induction} equations, the total transport of momentum expressed by the conserved angular velocity current $J^\omega$ is the sum of the contribution of Reynolds, Maxwell and viscous stresses \eqref{eq:stress}. At sufficiently large $Re$, the viscous contribution is confined to thin boundary layers attached to the cylinders. Because of the no-slip and insulating boundary conditions, at the cylinders there is only viscous transport (quantified by the torque $G$). As we are interested in the high $Re$ limit, in this section we analyze only the Maxwell and Reynolds contributions, which is consistent with the analysis of $G/G_\text{lam}-1$ presented in the previous section, \AG{and the characterization of the linear eigenmodes shown in Fig.~\ref{fig:linear_stress}}.

\begin{figure}
\centering
 \includegraphics[width=0.9\linewidth]{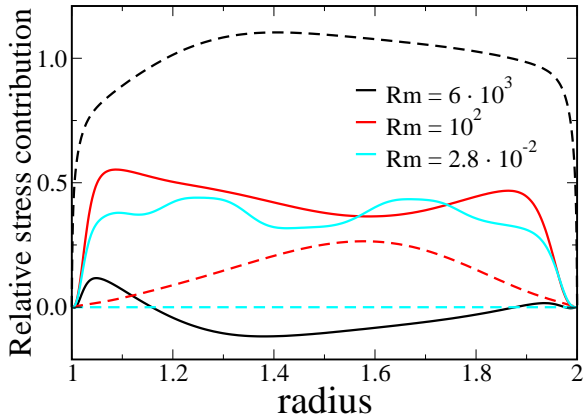}  
 \caption{\small Maxwell (dashed lines) and Reynolds (solid lines) stresses along the radius, normalized by the total angular velocity current $J^\omega$ at $\mu=0.26$.  Three cases are shown: $[Pm, Re] = [1.4\cdot 10^{-6}, 2\cdot 10^4]$ ($Rm=2.8\cdot 10^{-2}$, cyan), $[10^{-2},10^4]$ ($Rm=10^2$, red) and $[1, 6\cdot 10^3]$ ($Rm=6\cdot 10^3$, black).} 
  \label{fig:stresses_radius}
\end{figure}

\begin{figure}
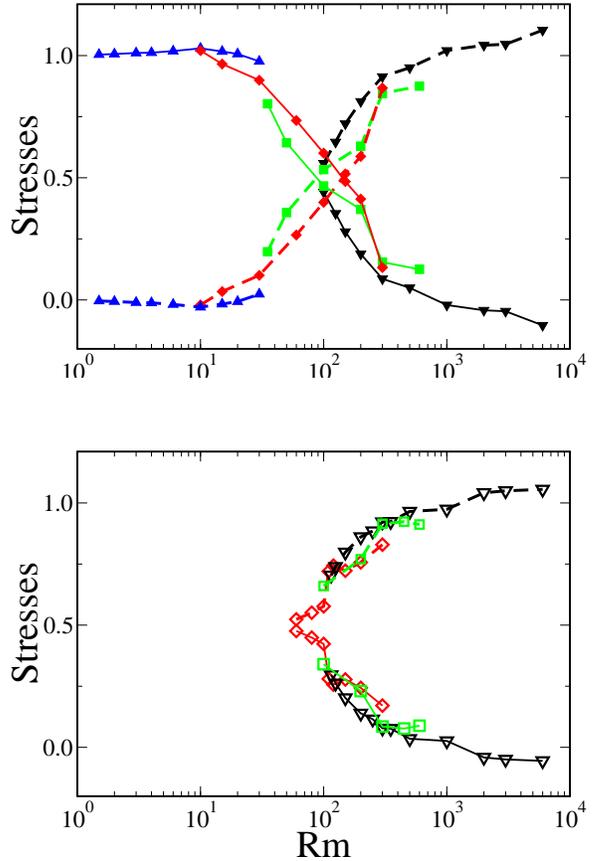

\centering   
 (a) \\
 \includegraphics[width=0.9\linewidth]{Max+Re_rayleigh_2.eps}  \\
     (b) \\
  \includegraphics[width=0.9\linewidth]{Max+Re_keplerian_2.eps}
 \caption{\small $Rm$-dependence of Maxwell and Reynolds stresses at the mid-gap for $\mu=0.26$  (a)  and $\mu=0.35$ (b). Different colors corresponds to data with different magnetic Prandtl number:  $Pm=10^{-3}$ (blue), $Pm=10^{-2}$ (red), $Pm=10^{-1}$ (green) and $Pm=1$ (black), as in Fig.~\ref{fig:torque}. Here solid and dashed lines denote the Reynolds and Maxwell stress contributions, respectively, normalized by their sum (i.e.~excluding the viscous contribution). } 
  \label{fig:stresses}
\end{figure}

Figure~\ref{fig:stresses_radius} presents the Reynolds stresses (solid) and Maxwell stresses (dashed) as functions of radius for three representative points in the parameter space. The stresses were normalized with the full azimuthal motion current $J^{\omega}$, which does not depend on $r$ according to \eqref{eq:stress}. Because of conservation of $J^{\omega}$, the viscous part can be obtained by subtracting the Maxwell and Reynolds contributions from 1. At $Pm=1.4 \cdot 10^{-6}, Re=2 \cdot 10^4$ ($Rm=2.8\cdot10^{-2}$,\,cyan) Maxwell stresses are negligible and up to 40\% of the angular momentum is transported by Reynolds stresses. Despite the large $Re$, viscous transport is still prevalent  and amounts to 60\% of the total. At $Pm=10^{-2}, Re=10^4$ ($Rm=10^2$,\,red) the Maxwell stresses amount to 20\% of the total transport in the middle of the gap, and Reynolds stresses contribute 40\%. For $Pm=1, Re=6 \cdot 10^3$ ($Rm=6\cdot10^3$,\,black) the Maxwell stresses dominate the transport in the center part of the domain, while Reynolds and viscous stresses are very small. Interestingly, the Reynolds stresses are negative in the center of the domain, corresponding to inward momentum transport due to velocity fluctuations. Here Maxwell stresses are larger than the total current in the middle part of the gap so that  $J^\omega$ remains conserved at each radial position. \AG{Note that the negative contribution of the Reynolds stress is also present in the linear eigenmode at $Pm=10^{-4}$ and $Rm=50$ shown in Fig.~\ref{fig:linear_stress}d}.

The magnetic Reynolds number $Rm= Re Pm$ increases in passing through the described points: ${Rm=[2.8 \cdot 10^{-2}; 10^2; 6 \cdot 10^3]}$, suggesting that the relative contribution of the stresses to the total current depends strongly on magnetic Reynolds number. \AG{This is again in line with the behavior of the linear eigenmodes discussed in section \S\ref{sec:linear}}. Figure~\ref{fig:stresses}a shows the contributions of Maxwell and Reynolds stresses, normalized by their sum, at the mid-gap as a function of magnetic Reynolds number. At low $Rm\lesssim10$ the contribution of Maxwell stresses is marginal, but thereafter   it begins to noticeably grow until it becomes equal to the Reynolds-stress contribution at $Rm\approx100$. If $Rm$ is increased further, Maxwell stresses dominate the turbulent angular momentum transport, and for $Rm\gtrsim10^3$ Reynolds stresses become negative and act as to counteract the outward transport by Maxwell stresses. Thus, $Rm =\mathcal{O}(100)$ marks \AG{the border between inertial-wave turbulence (excited by the imposed magnetic field) and turbulence arising from magnetocoriolis waves}, i.e.~the usual MRI for which magnetic stresses prevail.  This result is in agreement with \AG{our eigenmode analysis} and the work of \citet{gellert2016nonaxisymmetric}, who compared magnetic and kinetic energies and found them equal at $Rm \sim 200$. They considered so-called Chandrasekhar states, where magnetic and velocity fields have the same radial profiles (unlike here) in the similar range of $Rm \in [10^{-3}, 10^{5}]$. 

\AG{The evolution of stresses with $Rm$ for the quasi-Keplerian case ($\mu=0.35$) is shown in Fig.~\ref{fig:stresses}b. Because the flow is only unstable for $Rm\gtrsim 50$, here Maxwell stresses dominate directly from onset. Thereafter, the behavior is identical to that for rotation close to the Rayleigh line. Returning to the torque scaling of the form $(G/G_\text{lam} -1)=a\,Re'$ shown in Fig.~\ref{fig:torque}c, one observation can be made. All data with pre-factor $a\propto\sqrt{Pm}$ is  for $Rm>50$ (magnetocoriolis wave), whereas the scaling with constant pre-factor $a$  is in the regime $Rm<50$ (magnetically excited inertial wave).  Further, the data for $\mu=0.26$ and $Pm=10^{-2}$, spanning $10<Rm<400$, appears to feature a transition from one type of scaling to the other as $Rm$ increases (see Fig.~\ref{fig:torque}a), which also corresponds to the change in dominant eigenmode (see Fig.~\ref{fig:param}). Hence we conclude the crossover in transport scaling occurs at $Rm=\mathcal{O}(100)$.}
  
\section{Estimation of $\alpha_\text{eff}$ from torque}

\AG{In Taylor--Couette flow, angular momentum transport is exactly quantified by the torque, yet in the context of accretion-disk theory it is customary to use the $\alpha_\text{eff}$-parameter to quantify transport.} In this section, we estimate $\alpha_\text{eff}$ from our torque data. Our simulations show that the turbulent part of the torque is proportional to modified Reynolds number:
\begin{equation}\label{eq:G_a_Re}
G/G_\text{lam}-1 \approx a\, Re',
\end{equation}
where
\begin{equation}
a \sim \left\{
                \begin{array}{lr}
                \text{const}\, &  Rm < \mathcal{O}(100),\\
                Pm^{0.5}\, & Rm>\mathcal{O}(100).
                \end{array}
              \right.
\end{equation}

After inserting \eqref{eq:G_a_Re} in equation (\ref{eq:alpha_fin})  the expression for $\alpha_\text{eff}$ reads as
\begin{equation}\label{eq:alpha_a}
\alpha_\text{eff}=\frac{1}{2 \pi |q|} \frac{\left( a Re' +1 \right) G_\text{lam}}{Re^2}
\end{equation}
Considering $G_\text{lam} \propto Re$ from \eqref{eq:Glam_nondim} and $Re' \approx Re$ we find that $\alpha_\text{eff}$ scales as:
\begin{equation}
 \alpha_\text{eff} \propto  a+ 1/Re',
\end{equation}
and when $Re \to \infty$:
\begin{equation}
 \alpha_\text{eff} \propto a.
\end{equation}
Thus, effective viscosity is independent of $Pm$ for $Pm \leq 10^{-3}$ and scales as $Pm^{0.5}$ for $Pm \geq 10^{-2}$.

\AG{For near Rayleigh-line rotation at }$Pm=1.4 \cdot 10^{-6}$ the turbulent  torque scales with  $a = 2.7 \cdot 10^{-5}$. Inserting $a$, $q=-1.94$ and $G_\text{lam} \approx 12 Re'$ from \eqref{eq:Glam_nondim}  into \eqref{eq:alpha_a} we get a lower bound for $\alpha_\text{eff}$ in the limit of $Re \to \infty$:
\begin{equation}
\alpha_\text{eff} \approx   2.6  \cdot 10^{-5}.
\end{equation}
corresponding to the inductionless case. At $Pm=1$ the highest value for $\alpha_\text{eff} \approx 3.4 \cdot 10^{-3}$ is attained. \AG{In the case of quasi-Keplerian rotation, $q=-1.48$,  equation  \eqref{eq:Glam_nondim} gives $G_\text{lam} \approx 11\, Re'$ and  $\alpha_\text{eff} $ increases by a factor of $1.2$ when compared to near Rayleigh-line rotation.}

\AG{Here we must caution that in Taylor--Couette flow, because of the presence of the solid cylinders bounding the fluid, turbulence modifies the mean velocity profile. Thus,  the parameter $q$ in equation \eqref{eq:alpha_a}, estimated from the mean turbulent velocity, will depart from ideal values and at first will typically grow with $Re$.  In our simulations, $q$ calculated in the middle of the gap between cylinders changes, but always remains negative. As turbulence becomes fully developed, $q$ appears to saturate at around $-1/2$, which is a (more) hydrodynamically stable velocity profile, unstable to MRI. This saturation  implies independence of  $\alpha_\text{eff}$ in $Re \to \infty$ regime. The key feature of the flow here is the proportionality to $Re$ of the outward turbulent transport of momentum; using the turbulent estimate $q=-1/2$ simply results in values of $\alpha_\text{eff}$ a factor of $3$ higher, as seen from equation \eqref{eq:alpha_fin}. Finally, we speculate that even if the mean flow profile were forced to remain unchanged (quasi-Keplerian), this scaling may remain unaffected, as recently observed in DNS of self-sustained quasi-Keplerian dynamos in Taylor--Couette flow \citep{guseva2017dynamo}. }

\section{Discussion}

\AG{We have performed a comprehensive study of the azimuthal MRI in Taylor--Couette flow for wide range of parameters $Pm \in [0, 1]$ and $Re$ up to $4 \cdot 10^4$. Two distinct velocity profiles were considered, namely quasi-Keplerian ($\mu=0.35$) and almost-constant specific angular momentum ($\mu=0.26$). Our linear stability analysis and direct numerical simulations highlight the relevance of the magnetic Reynolds number $Rm$ in determining the radial transport of angular momentum. For $Rm>\mathcal{O}(100)$, regardless of rotation profile, the flow is unstable to the usual MRI.  Transport is governed by Maxwell stresses  and scales as $\sqrt{Pm}Re^2$, so that $\alpha_\text{eff}\propto \sqrt{Pm}$, consistent with  \citet{rudiger2015angular}. At $Pm=1$ we found $\alpha_\text{eff}>10^{-3}$ at $Pm=1$. Hence in highly ionized disks or disk regions, the AMRI may be a vigorous source of angular momentum transport. At low $Rm<\mathcal{O}(100)$, instability is found only for steep profiles very close to the Rayleigh-line. Here the flow is unstable to the inductionless MRI  for hydrodynamic Reynolds number $Re \gtrsim1000$, transport is governed by Reynolds stresses, scales as $Re^2$ and is weak with $\alpha_\text{eff}=\mathcal{O}(10^{-5})$.}

\AG{The ratio of Maxwell to Reynolds stresses is solely determined by $Rm$ and increases from zero to one as $Rm$ is increased. This is in agreement with \citet{meheut2015angular}, who performed shearing-box simulations of MRI turbulence at two magnetic Reynolds numbers $Rm=400$  and $2600$, while varying either $Pm$. At each $Rm$ they found a constant, but different, ratio of Maxwell to Reynolds stress both for azimuthal and axial magnetic fields, with Maxwell stresses growing with $Rm$.}

\AG{Our results are in line with the linear analysis of \cite{Kirillov2010} for the MRI with imposed helical magnetic fields. They identified the inductionless instability close to the Rayleigh line as a magnetically destabilized inertial wave, whereas the usual MRI can be interpreted as arising from an unstable magnetocoriolis wave \citep{nornberg2010observation}. In addition, \cite{Kirillov2010} showed that the transfer of instability between the two modes was continuous. Our data support also a similar scenario for the AMRI: for $\mu=0.26$ and $Pm=0.01$ a continuous transition between the two types of turbulent flow can be observed as $Rm$ increases. Thus we suggest that the dependence of scaling type on $Rm$ shall also apply to MRI in the presence of helical magnetic fields. Future  liquid metal experiments planned by \citet{stefani2017dresdyn}, aiming at  $Rm>\mathcal{O}(10)$, should confirm the crossover between the two flavors of MRI  and transport scalings shown here}. 
 
\acknowledgements
This research was supported in part by the National Science Foundation under Grant No. NSF PHY-1125915. Support from the Deutsche Forschungsgemeinschaft (grant number AV 120/1-1) and computing time from the Regional Computing Center of Erlangen (RRZE) and the North-German Supercomputing Alliance (HLRN) are greatfully acknowledged.

\bibliographystyle{aasjournal}
\bibliography{Bibliography}

\end{document}